\documentclass[aps,pra,preprintnumbers,groupedaddress,nofootinbib,showpacs,twocolumn]{revtex4}

\usepackage{graphics}
\usepackage{picins}
\usepackage[colorlinks=true]{hyperref}
\usepackage{color}
\begin{document}

\title{Spin Hall effect on a noncommutative space}
\author{Kai Ma}
\author{Sayipjamal Dulat}

\email{dulat98@yahoo.com}

\affiliation{School of Physics Science and Technology, Xinjiang
University, Urumqi, 830046, P. R. China}

\begin{abstract}
\noindent We study the spin-orbital interaction and
the  spin Hall effect of an electron moving on a noncommutative
space under the influence of a vector potential $\vec{A}$. On a
noncommutative space we find that the commutator between the vector
potential $\vec{A}$ and the electric potential $V_1(\vec r)$ of
lattice  induces a new term which can be treated as an effective
electric field, and the spin Hall conductivity obtains some
correction. On a noncommutative space the spin current and spin Hall
conductivity have distinct values in different direction, and
depends explicitly on the noncommutative parameter. Once this spin
Hall conductivity in different direction can be measured experimentally with a high
accuracy, the data can be used to impose bounds on the value of the
space noncommutativity parameter. We have also defined a
new parameter $\varsigma=\rho\theta$ ($\rho$ is the electron
concentration, $\theta$ is the noncommutativity parameter) which can be
measured experimentally.  Our approach is based on the
Foldy-Wouthuysen transformation which give a general Hamiltonian of
a non-relativistic electron moving on a noncommutative space.
\end{abstract}

\pacs{71.70.Ej, 02.40.Gh, 03.65.-w}

\maketitle

\section{Introduction}\label{intro}
The approach to noncommutative quantum field theory based on star
products and Seiberg-Witten maps allows for the generalization of
the standard model of particle physics to the case of
noncommutative space-time. Since noncommutative quantum field
theory may solve the puzzles of the Standard model, there are a lot
of papers concerning the quantum field theory on a noncommutative
space-time \cite{NCQEDGG}-\cite{NCQED}. Apart from these studies,
many articles have been devoted to the study of various aspects of
quantum mechanics (QM) on a noncommutative space (NCS) and a
noncommutative phase space(NCPS), because the main goal of the
noncommutative quantum mechanics (NCQM)
is to find a measurable spatial noncommutativity effects. For
example, the authors in Ref.~\cite{NCPH,NCPRH} applied
NCQM to the Hydrogen atom and
calculated the corrections to the Lamb shift. They find the
constraint on $\theta$ is $1/\sqrt\theta \geq 10^{2}GeV$.
Ref.~\cite{Haghighat} provided a
constraint: $1/\sqrt\theta \geq 3GeV$ by studying the transitions in
the Helium atom.  A possibility of testing spatial noncommutativity
via cold Rydberg atoms is suggested in Ref.\cite{NCPR}. The authors
in Refs.~\cite{chaichain1}-\cite{NCABLS} have studied the
Aharonov-Bohm phase on a NCS and a NCPS. A lower bound
$1/\sqrt\theta \geq 10^{-6}GeV$ for the space  noncommutativity
parameter is obtained \cite{chaichain1}. The Aharonov-Casher phase
for a spin-1/2 and a spin-1 particle on a NCS and a NCPS has been
studied in Refs.~\cite{mirza1}-\cite{mirza2}. A limit $1/\sqrt\theta
\geq 10^{-7}GeV$ for the space noncommutativity parameter is
obtained \cite{mirza1}. The noncommutative quantum Hall effect has
been studied in Refs. \cite{harms}-\cite{saha}, and
Ref.~\cite{harms} found a lower limit of $1/\sqrt\theta \geq 10GeV$
on the noncommutativity parameter.  Ref. \cite{newsh4} discussed
noncommutative spin Hall effect (SHE) through a semiclassical
constrained Hamiltonian, and obtained interesting results.
One can use the data on the the spin Hall measurements\cite{SHEexperimentaldata} to impose some bounds on the value of the noncommutativity parameter $1/\sqrt\theta$, and  we find that the limit on
$1/\sqrt\theta$ is weaker by $thirteen$ orders of magnitude than the one
imposed by the quantum Hall effect \cite{harms}. However we propose a
stronger limit on $1/\sqrt\theta$ by measuring the spin Hall
conductivity in different directions. The SHE is
predicted by M.I. Dyakonov and V. I. Perel in 1971\cite{dya}. This
effect, which occur as a result of the spin-orbit coupling
between electrons and impurities, is called extrinsic.  Conversely,
there are intrinsic forms of the SHE, which is caused by spin-orbit
coupling in the band structure of the semiconductor and survives in
the limit of zero disorder, which have become an active field of
research in recent years \cite{SHall1}-\cite{SOI}.

In this paper we focus on the intrinsic SHE on a NCS. To begin, we must define what we mean by ``noncommutative space". NCS is a deformation of ordinary space in
which the space coordinate operators $\hat x_i$ satisfy the
following relation,
\begin{equation}\label{NCRSS}
    [\hat{x}_i,\hat{x}_j]=i\theta_{ij},
    ~~~~~~~~~~~~~~~~~~~~~
\end{equation}
\begin{equation}\label{NCRSP}
    [\hat{x}_{i},\hat{p}_{j}]=i\hbar\delta_{ij},
    ~~~~~~~~~~~~~~~~~~~
\end{equation}
\begin{equation}\label{NCRPP}
    [\hat{p}_i,\hat{p}_j]=0, ~~~i,j=1,2,\cdots,n
\end{equation}
 $\theta_{ij}$ is totally antisymmetric real tensor which represent the
noncommutativity of  the space. In addition, the product between
the external fields on a NCS is,
\begin{equation}\label{SP}
    \hat{f}(\hat{\vec{x}})\hat{g}(\hat{\vec{x}})
    \equiv f(\vec{x})\ast g(\vec{x})
    =\exp{\left[\frac{i\theta_{ij}}{2}
    \partial_{x_{i}}\partial_{y_{j}}\right]}
    f(\vec{x})g(\vec{y})|_{\vec{x}=\vec{y}},
\end{equation}
where $f(\vec{x})$ and $g(\vec{x})$ are two arbitrary infinitely
differentiable functions on a commutative $R^{3+1}$ space, and
$\hat{f}(\hat{\vec{x}})$ and $\hat{g}(\hat{\vec{x}})$ are the
corresponding functions on a NCS. In NCQM one
replaces the position operator $\hat x_i$ with the $\theta$-deformed
$$
\hat x_i = x_i +\frac{1}{\hbar}\theta_{ij}p_j\;,\hspace{0.6cm}
i,j=1,2\cdots, n
$$
and then applies the usual rules of QM. We use the
following notation for the commutator of two arbitrary physical
observable operators $\hat A$ and $\hat B$ in NCQM,
\begin{equation}\label{NCC}
    [\hat A,\hat B]= \hat A \hat B -\hat B\hat A=
    [A,B]_{\ast}=A\ast B-B\ast A.
\end{equation}
Note that on a NCS
$\hat{\vec A }\times \hat{ \vec A}\neq0$, this can be seen clearly
via,
\begin{equation}\label{NCCP}
   \hat{\vec A}\times\hat{\vec A}
   =
    \vec A \times_{\ast}\vec A
    =A_{i}\ast A_{j}
    \epsilon_{ijk}\vec{e}_{k}\neq A_{j}
    \ast A_{i}\epsilon_{ijk}\vec{e}_{k}
    \neq0.
\end{equation}
where  the $\vec A \times_{\ast}\vec A=\hat{\vec A}\times\hat{\vec
A}$ indicates cross product on a NCS. Note that one
of the most interesting things in noncommutative quantum field
theory is that even the $U(1)$ gauge group has non-Abelian like
characteristics such as self-interactions.

 This paper is organized as follows: in the
next section we will provide the Foldy-Wouthuysen Hamiltonian of a
non-relativistic spin-1/2 particle moving on a NCS
under the influence of some external electromagnetic fields.  In
section 3, we will calculate the spin-dependent electric current and
the spin Hall conductivity. The conclusion is given in the final
section.

\section{Spin-Orbital Interaction}\label{SOInter}

In this section we will present the general form of Foldy-Wouthuysen
Hamiltonian $H_{FW}$ of an electron in the presence of external
fields on a NCS. As we know in the Dirac
representation the wave function of a free spin-half particle is
generally determined by four component spinors. And the Dirac
equation has positive and negative total energy solutions. In the
non-relativistic limit, we are usually only interested in the
positive energy part of the spectrum that describes electrons
(rather than positrons). The Foldy-Wouthuysen transformation
has been proposed in Ref.\cite{wf}. Its main goals are
transformation of Dirac Hamiltonians to the block-diagonal form, (it
separates distinct components of relativistic spinor wave function
and energy spectrum), and establishment of connection between the
relativistic QM and the QM as well as the
classical physics.

Let us start with the Dirac Hamiltonian for a particle with charge
$e$ interacting with an external electromagnetic field,
\begin{equation}\label{HD}
H=c\vec{\alpha}\cdot(\vec{p}-\frac{e}{c} \vec{A})+ \beta mc^2 +
eV(\vec{r}),
\end{equation}
where $\beta=\gamma_0$ and $\alpha_i =\gamma_0\gamma_i$ are the
Dirac matrices; $\vec p=-i\hbar\vec\bigtriangledown$ is the canonical
momentum operators; $\vec A$ is the vector potential.  The
non-relativistic limit of the Dirac equation (\ref{HD}) is reached
by  a Foldy-Wouthuysen unitary transformation, which block
diagonalizes the Dirac Hamiltonian by separating the positive and
negative energy part of its spectrum.
By a unitary transformation \cite{RQMG} the transformed Foldy-
Wouthuysen Hamiltonian is given by,
\begin{equation}\label{CSFWH}
    H_{FW}=\beta\bigg(mc^2+\frac{O^2}{2mc^2}
    \bigg)+\epsilon-\frac{1}{8m^2c^4}[O,[O,
    \epsilon]],
\end{equation}
where,
\begin{equation}\label{EEM}
    O=c\vec{\alpha}\cdot(\vec{p}-\frac{e}{c}
    \vec{A}),~~~~\epsilon=eV(\vec{r}).
\end{equation}
On a NCS the Foldy-Wouthuysen Hamiltonian is,
\begin{equation}\label{FWH}
    \hat{H}_{FW}=\beta\bigg(mc^2+\frac
    {\hat{O}^2}{2mc^2}\bigg)+\hat{\epsilon}
    -\frac{1}{8m^2c^4}[\hat{O},[\hat{O},
    \hat{\epsilon}]].
\end{equation}
We now calculate the various terms of $\hat{H}_{FW}$ explicitly. For
the second term we have,
\begin{eqnarray}\label{O2}
    \hat{O}^2&=&O\ast O=
    c^2(\vec{p}-\frac{e}{c}
    \vec{A})\ast(\vec{p}-\frac{e}{c}
    \vec{A})\nonumber \\
    &&-ec\hbar\vec{\Sigma}
    \cdot\vec{B}+ie^2\vec{\Sigma}
    \cdot(\vec{A}\times_{\ast}\vec{A})\;,
\end{eqnarray}
here we have used  the following relations for two arbitrary vectors
$\vec A$ and $\vec B$,
$$
(\vec\alpha\cdot\vec A)(\vec\alpha\cdot\vec B)=\vec A\cdot\vec B
+i\vec\Sigma\cdot(\vec A \times \vec B)\;,
$$
and,
$$curl(f\vec A)=f~curl\vec A +grad\;f\times \vec A\;.$$
as well as equation(\ref{NCCP}). Next we look at the commutator
$[\hat{O},\hat{\epsilon}]$,
\begin{eqnarray}\label{Oe}
    [\hat{O},\hat{\epsilon}]&=&
    [O,\epsilon]_{\ast}\nonumber\\
    &=&-ice\hbar\vec{\alpha}\cdot
    \vec{\nabla}V(\vec r)-e^2\vec{\alpha}
    \cdot[\vec{A},V]_{\ast}.\nonumber
\end{eqnarray}
Furthermore, we can calculate the commutator,
\begin{eqnarray}\label{OOe}
    [\hat{O},[\hat{O},\hat{\epsilon}]]&=&
    [O,[O,\epsilon]_{\ast}]_{\ast}
    \nonumber\\ &=&
    ec^2\hbar\vec{\nabla}\cdot
    \vec{E}+iec^2\hbar^2\vec
    {\Sigma}\cdot(\vec{\nabla}\times\vec{E})
    \nonumber\\
    &&+2ec^2\hbar\vec{\Sigma}\cdot
    (\vec{E}\times\vec{p})\nonumber\\
    &&-2ie^2c\vec{\Sigma}\cdot(
    [\vec{A},V]_{\ast}\times\vec{p}).
\end{eqnarray}
In the calculations above we have neglected the higher order terms
in $\theta$ such as $[\vec{A},[\vec{A},V]_{\ast}]_{\ast}$ and
$\vec{\nabla}\cdot[\vec{A},V]_{\ast}$. Adding the various terms we
have,
\begin{eqnarray}\label{FWHE}
    \hat{H}_{FW}&=&\beta\bigg( mc^2+\frac{1}{2m}
    (\vec{p}-e\vec{A}/c)\ast
    (\vec{p}-e\vec{A}/c)\bigg)\nonumber\\
    &&+ eV(\vec{r})-\frac{e\hbar\beta}{2mc}\vec{\Sigma}
    \cdot\vec{B}-\frac{ie\hbar^2}{8m^2c^2}
    \vec{\Sigma}\cdot(\vec{\nabla}
    \times\vec{E})\nonumber\\
    &&-\frac{e\hbar}{4m^2c^2}\vec{\Sigma}\cdot
    (\vec{E}\times\vec{p})-\frac{e\hbar^2}
    {8m^2c^2}\vec{\nabla}\cdot\vec{E}\nonumber\\
    &&-\frac{e\hbar}{4m^2c^2}
    \vec{\Sigma}\cdot(\hat{\vec E}
    \times\vec{p})
    -\frac{e\hbar\beta}{2mc}\vec{\Sigma}
    \cdot\hat{\vec B }\;,
\end{eqnarray}
where $\hat{\vec E }=-\frac{ie}{c\hbar}[\vec{A},V]_{\ast}$ and $\hat
{\vec B}=-\frac{ie}{c\hbar}\vec{A} \times_{\ast}\vec{A}$. The FW
Hamiltonian (\ref{FWHE}) describes the dynamics of an electron, but
the model will also equally apply to a positron ( which corresponds
to a hole in solid). Now we discuss the individual terms of
(\ref{FWHE}). The terms in the first parenthesis result from the
expansion of $[(\vec{p}-e/c\vec{A})^2+m]^{1/2}$ and describe the
relativistic mass increases, and we will neglect the corresponding
noncommutative corrections in the next section. The second term
describes the electrostatic energy. The third term is a magnetic
dipole energy which induces Zeeman effect. The next two terms
contain the spin-orbital interaction (Rashba spin-orbital coupling
for the constant electric field which is along the $ \vec{z}$
direction $\vec{E}\sim \vec{k}$ (unite vector)). This can be seen
clearly under the assumption of a spherically symmetric potential
with $\vec{\nabla}\times\vec{E}=0$, then we have,
\begin{equation}\label{SOU}
    \vec{\sigma}\cdot(\vec{E}\times\vec{p})
    =-\frac{1}{r}\frac{\partial V}{\partial
    r}(\vec{\sigma}\cdot\vec{L}).
\end{equation}
The last two terms are the noncommutative  corrections of the
electron moving on a NCS. For the term
$\vec{\Sigma}\cdot(\hat{\vec E}\times\vec{p})$, explicitly, it is an
effective spin-orbital interaction and the corresponding effective
electric field is $\hat {\vec E}=-ie/(c\hbar)
[\vec{A},V(\vec{r})]_{\ast}$. The term $\vec{\Sigma}\cdot \hat {\vec
B }$ is an effective Zeeman interaction, and the effective magnetic
field is,
\begin{equation}\label{NCEB}
   \hat{ \vec B }=-\frac{ie}{c\hbar}\vec{A}
    \times_{\ast}\vec{A}=-\frac{ie}
    {c\hbar}(x_2\ast x_1-x_1
    \ast x_2)\vec{z}=\frac{e\theta}
    {c\hbar}\vec{z}.
\end{equation}
Note that the important physical quantity for the dynamics of the
charged particles is the vector potential $\vec A$, and one can
understand this via Aharonov-Bohm effect. Thus in our discussion we
will impose a particular situation where $\vec B=\nabla\times\vec A$
and the Coulomb gauge $\vec{\nabla}\cdot\vec A=0$, where  $\vec
A=(x_2,x_1,0)$.

\section{Spin Hall Effect}\label{NCSHalleff}

In this section we will calculate the spin-depended electric
current. This is performed by incorporating spin and spin-orbital
interaction into the dynamics of charge carriers as in
Ref.~\cite{SDECMa}. This allows one to obtain
universal expression for the spin Hall conductivity on a
NCS. We must notice that for a macroscopic system
the total electric potential $V(\vec{r})$ is the sum of external
electric potential $V_0(\vec{r})$ and the lattice electric potential
$V_1(\vec{r})$. Collecting the dynamical terms and spin-orbital
coupling terms in the Hamiltonian (\ref{FWHE}) we have,
\begin{eqnarray}\label{SOIH}
    \hat{{\mathcal {H}}}&=&\frac {\vec p^2}{2m}
    +\frac{e\hbar}{4m^2c^2}\vec{\sigma}
    \cdot\bigg[\bigg(\vec{\nabla}V(\vec{r})-\frac{ie}
    {c\hbar}[\vec{A},V]_{\ast}\bigg)
    \times\vec{p}\bigg]\nonumber
    \\&&+eV(\vec{r})\nonumber\\
    &\equiv&\frac{\vec{p}^2}{2m}
    -\frac{e\hbar}{4m^2c^2}\vec{\sigma}
    \cdot\bigg(\hat{\vec{\mathcal E}}
    \times\vec{p}\bigg)+eV(\vec{r}),
\end{eqnarray}
where $\hat{\vec{\mathcal E}}=-\vec{\nabla}V(\vec{r})
+\frac{ie}{c\hbar}[\vec{A},V]_{\ast}$. By now we have derived the
general formalism for spin-orbital interaction on a NCS. Heisenberg algebra for canonically conjugated variables
$\vec{p}$ and $\vec{r}$ on a NCS is
\begin{eqnarray}\label{HAC}
    \dot{\hat{ \vec r}}&=&\frac{1}
    {i\hbar}[\vec{r},\hat{{\mathcal
    {H}}}]=\frac{\vec{p}}{m}+\frac{e
    \hbar}{4m^2c^2}\vec{\sigma}\times
    \vec{\nabla}V(\vec{r})\nonumber
    \\&&-\frac{ie^2}
    {4m^2c^3}\vec{\sigma}\times
    [\vec{A},V]_{\ast} \;,\\\label{HAP1}
    \dot{\hat{\vec p}}&=&\frac{1}{i\hbar}
    [\vec{p},\hat{{\mathcal {H}}}]=
    -e\vec{\nabla}V(\vec{r})\nonumber
    \\&&-\frac{e\hbar}
    {4m^2c^2}\vec{\nabla}\bigg[\bigg(
    \vec{\sigma}\times\vec{\nabla}V(\vec{r})
    \bigg)\cdot\vec{p}\bigg]\nonumber
    \\&&+\frac{ie^2}
    {4m^2c^3}\vec{\nabla}\bigg[\bigg(
    \vec{\sigma}\times[\vec{A},V]_{\ast}
    \bigg)\cdot\vec{p}\bigg]\;.
\end{eqnarray}
From (\ref{HAC}) we have,
\begin{eqnarray}\label{PE}
    \hat{\vec{p}} &=&m\dot{\hat{\vec{r}}}-\frac
    {e\hbar}{4mc^2}\vec{\sigma}\times
    \vec{\nabla}V(\vec{r})\nonumber
    \\&&+\frac{ie^2}
    {4m^2c^3}\vec{\sigma}\times
    [\vec{A},V]_{\ast},\\\label{AE1}
    \hat {\dot {\vec p}}&=& m\ddot
    {\hat{\vec{r}}}-\frac{e\hbar}
    {4mc^2}\bigg(\dot{\vec{r}}\cdot
    \vec{\nabla}\bigg)\bigg(\vec{\sigma}
    \times\vec{\nabla}V(\vec{r})\bigg)
    \nonumber\\&&+\frac{ie^2}
    {4m^2c^3}\bigg(\dot{\vec{r}}\cdot
    \vec{\nabla}\bigg)\bigg(\vec{\sigma}
    \times[\vec{A},V]_{\ast}\bigg).
\end{eqnarray}
The second term in (\ref{PE}) is the cross product of the electron
magnetic moment and the effective  electric field on a NCS.
Substitution of (\ref{PE}) and (\ref{AE1}) into (\ref{HAP1}) yields
the following form of the  Newton's second law
 for charge carriers on a NCS
\begin{equation}\label{NFR}
    m\ddot{\hat{\vec{r}}}=
    \hat{\vec F}(\vec{\sigma})
    =\vec{F}(\vec{\sigma})+
    \vec{F}_{\theta}(\vec{\sigma}),
\end{equation}
where  the spin-dependent force $\hat{\vec F}(\vec{\sigma})$ on a
NCS is given by,
\begin{equation}\label{NFE}
    \vec{F}(\vec{\sigma})=-\frac{e\hbar}
    {4mc^2}\dot{\vec{r}}\times\bigg[
    \vec{\nabla}\times\bigg(\vec{\sigma}
    \times\vec{\nabla}V(\vec{r})\bigg)
    \bigg]-e\vec{\nabla}V(\vec{r}),
\end{equation}
\begin{equation}\label{NFEtheta}
    \vec{F}_{\theta}(\vec{\sigma})=\frac{ie^2}
    {4mc^3}\dot{\vec{r}}\times\bigg[
    \vec{\nabla}\times\bigg(\vec{\sigma}
    \times[\vec{A},V]_{\ast}\bigg)
    \bigg].~~~~~~~~~
\end{equation}
Here we neglected the  terms proportional to  $1/c^4$. Note that the
force in (\ref{NFR}) is equivalent to the Lorentz force,
\begin{equation}\label{LFE}
    \hat{\vec F}(\vec{\sigma})=\frac{e}
    {c}\bigg(\dot{\vec{r}}
    \times\hat{\vec B}(\vec{\sigma})
    \bigg)-e\vec{\nabla}V(\vec{r}).
\end{equation}
which acts on a particle of charge $Q=e$ in the electric field
$\vec{E}=-\vec{\nabla}V(\vec{r})$ and magnetic field,

\begin{equation}\label{LFE}
    \hat{ \vec B}(\vec{\sigma})=
    \vec{\nabla}\times\hat{\vec A}(\vec{\sigma})=
    \vec{\nabla}
    \times[\vec{A}(\vec{\sigma})+
    \vec{A}_{\theta}(\vec{\sigma})]~,
\end{equation}
where
 \begin{eqnarray}
 \vec A(\vec{\sigma})
    &=&-\frac{\hbar}{4mc}\vec{\sigma}\times
    \vec{\nabla}V(\vec{r})~,\\
    \vec A_{\theta}(\vec{\sigma})
    &=&\frac{ie}{4mc^2}\vec{\sigma}\times
    [\vec{A},V]_{\ast}~.
\end{eqnarray}
The Hamiltonian (\ref{SOIH}) can be written as,
\begin{equation}\label{RSOIH}
    \hat{{\mathcal {H}}}=\frac{1}{2m}\bigg(
    \vec{p}-\frac{e}{c}\hat{\vec A}(\vec{\sigma})\;.
    \bigg)^2
\end{equation}

To solve the equation (\ref{NFR}), the velocity relaxation time
$\tau$ must be given experimentally. We assume that to the first
approximation the velocity relaxation time $\tau$ of charge carriers
is independent of $\vec{\sigma}$. Because of relative smallness of
the spin-dependent force, we can treat $\hat{\vec F}(\vec \sigma)$
in (\ref{NFR}) as a perturbation. The solution of (\ref{NFR}) can be
written in the form
$\dot{\hat{\vec{r}}}=\dot{\vec{r}}+\dot{\vec{r}}_{\theta}$, where
$\dot{\vec{r}}$ is the $\theta$-independent solution that is given
in Ref.\cite{SDECMa}, and $\dot{\vec{r}}_{\theta}$ is a small
$\theta$-dependent part of the velocity which also can be obtained
perturbatively. In the presence of a constant external electric
field $\vec{E}=-\vec{\nabla}V_0(\vec{r})$, from (\ref{NFR}),
(\ref{NFE}) and (\ref{NFEtheta}) we obtain,
\begin{equation}\label{FA}
    \langle\dot{\vec{r}}\rangle=
    \frac{e\tau}{m}\vec{E}-\frac
    {\hbar e^2\tau^2}{4m^3c^2}
    \vec{E}\times\langle\vec{\nabla}
    \times[\vec{\sigma}\times
    \vec{\nabla}V(\vec{r})]\rangle\;,
\end{equation}
\begin{equation}\label{SA}
    \langle\dot{\vec{r}}_{\theta}\rangle
    =\frac{ie^3\tau^2}{4m^3c^3}
    \vec{E}\times\langle\vec{\nabla}
    \times(\vec{\sigma}\times
    [\vec{A},V]_{\ast})\rangle \;.
    ~~~~~~~~
\end{equation}
Using our simple choice for vector potential $\vec{A}=(x_2,x_1,0)$,
we can obtain,
\begin{equation}\label{EEE}
    [\vec{A},V]_{\ast}=2i\theta\stackrel
    {\Rightarrow}{\eta}\cdot\vec
    \nabla V(\vec{r}),
\end{equation}
where $\stackrel{\Rightarrow}{\eta}$ is a tensor with the following
components $\eta_{11}=1, \eta_{22}=-1, \eta_{12}=\eta_{21}=0$. The
right-hand side of (\ref{FA}) and (\ref{SA}) contains the volume
average of electrostatic crystal potential
$\partial_i\partial_jV_1(\vec{r})$. The SHE in a cubic lattice on a
commutative space has been studied in Ref.~\cite{SDECMa}. In the
following we will study the noncommutative SHE in a cubic lattice.
On a commutative space, for a cubic lattice, the only invariant
permitted by symmetry is,
\begin{equation}\label{CP}
    \langle\frac{\partial^2V_1(\vec{r})}
    {\partial r_{i}\partial r_{j}}\rangle
    =\chi\delta_{ij},
\end{equation}
where $\chi$ is a constant which have been determined in
Ref.~\cite{SDECMa}. However, on a NCS, the spacial
symmetry is twisted by the noncommutativity of space coordinates
through the commutator between vector potential and electric
potential. This correction is represented by formalism (\ref{EEE})
up to first order of $\theta$. Here, we are only interested in the
first order correction. With the help of (\ref{EEE}) and (\ref{CP})
we obtain,
\begin{equation}\label{FAVA}
    \langle\dot{\vec{r}}_{\theta}\rangle=
    \frac{e^3\tau^2\chi\theta}{m^3c^3}
    (\stackrel{\Rightarrow}{\eta}\cdot
    \vec{\sigma}\times\vec{E}).
\end{equation}
The density matrix of the charge carriers in the spin space can be
written as,
\begin{equation}\label{SDM}
    \rho^{s}=\frac{1}{2}\rho(1+\vec{\lambda}
    \cdot\vec{\sigma}),
\end{equation}
where $\rho$ is the total concentration of charges carrying the
electric current; $\vec{\lambda}$ is  the vector of spin
polarization of the electron fluid. The $\theta$-independent and
$\theta$-dependent currents are,
\begin{eqnarray}\label{EJ}
    &&\vec{j}=e\langle\rho^{s}
    \dot{\vec{r}}\rangle\equiv
    \vec{j}^{0}+\vec{j}^{s}
    (\vec{\sigma}),\nonumber\\
    &&\vec{j}^{0}=\sigma_{H}\vec{E},
    ~~~\vec{j}^{s}(\vec{\sigma})
    =\sigma_{H}^{s}(\vec{\lambda}
    \times\vec{E}),~~~~~~~~~~~~
\end{eqnarray}
\begin{equation}\label{SJ}
    \vec{j}^{s}_{\theta}(\vec{\sigma})=e\langle
    \rho^{s}\dot{\vec{r}}_{\theta}\rangle=
    \sigma_{H_{\theta}}^{s}(\stackrel
    {\Rightarrow}{\eta}\cdot\vec{\lambda}
    \times\vec{E}),~~~~
\end{equation}
where the corresponding Hall conductivities are given by,
\begin{equation}\label{HC}
    \sigma_{H}=\frac{e^2\tau\rho}
    {m},~~~~~~
\end{equation}
\begin{equation}\label{SHConduct}
    \sigma_{H}^{s}=\frac{\hbar e^3\tau^2\rho
    \chi}{2m^3c^2},~
\end{equation}
\begin{equation}\label{NCSHConduct}
    \sigma_{H_{\theta}}^{s}=\frac{e^4\tau^2\chi
    \varsigma}{m^3c^3},
\end{equation}
where $\varsigma=\rho\theta$, and  $\rho\sim10^{22}cm^{-3}$.
Thus we see that the noncommutativity parameter $\theta$ makes an important contribution to the spin Hall conductivity. Explicitly, its contribution to the spin current
$\hat{\vec{j}}^{s}=\vec{j}^{s}+\vec{j}^{s}_{\theta}$ and
spin Hall conductivity
$\hat{\sigma}_{H}^{s}=\sigma_{H}^{s}+\sigma_{H_{\theta}}^{s}$ is order of $e\theta/(c\hbar)$.
Furthermore, if the external electric field $\vec E$ is along the
$z$-axis,  the  components of the spin current can be written as,
\begin{equation}\label{Spincx}
    \hat{j}^{s}_{x}(\vec{\sigma})=
    (\sigma_{H}^{s}-\sigma_{H_{\theta}}^{s})E,
\end{equation}
\begin{equation}\label{Spincy}
    \hat{j}^{s}_{y}(\vec{\sigma})=
    -(\sigma_{H}^{s}+\sigma_{H_{\theta}}^{s})E,
\end{equation}
where $E$ is the absolute value of $\vec E$.  Note that on a
NCS, the system has distinct transport behavior in
different directions, namely, the absolute spin Hall conductivity
has different values in different directions, and the separation is
$2\sigma_{H_{\theta}}^{s}$. If we can measure it with sufficient
precision, then we would have a clear evidence
for the space noncommutativity.  If $\theta=0$, our result agrees
with that of Ref.~\cite{SDECMa}. Note also that since the noncommutativity of space
breaks rotational symmetry, the non-trivial
noncommutative correction terms of the spin Hall conductivity have a preferred direction.

\section{Conclusion}

The origin of SHE is the spin-orbit interaction. By using the
Foldy-Wouthuysen transformation, which give a general Hamiltonian of
a non-relativistic moving Dirac particle, we derived the general
form of the effective spin-orbital coupling on a NCS, in which the upper and lower two components of the electrons
wave function are separated completely. Thus our formalisms are
applicable both for electrons and positrons. In our investigation of
the spin Hall effect, we consider a particular situation, namely,
the electron is influenced by the vector potential $\vec{A}$ but the
magnetic field $\vec{B}$ is absent. With this choice the first order
correction of (\ref{EEE}) is obtained in a natural way. After
solving the ``Newton equation" perturbatively in the framework of
NCQM, we derived the spin-depended
electric current whose expectation value gives the spin Hall effect
and spin Hall conductivity on a NCS. We find that
on a NCS the commutator between vector potential
$\vec{A}$ and electric potential of lattice $V_1(\vec{r})$ induces a
new term which can be treated as an effective electric field, and
the spin Hall conductivity is corrected accordingly. In the presence
of the NCS the spin Hall conductivity takes
different values in different direction, and depends explicitly on
the noncommutativity parameter.
In addition the noncommutative correction ratio is,
\begin{equation}\label{nccorrectionratio}
    R=\sigma^{s}_{H_\theta}/\sigma^{s}_{H}=2e\theta/(c\hbar).
\end{equation}
By using experimental data \cite{SHEexperimentaldata} of spin
Hall conductivity we can impose a lower limit of $1/\sqrt\theta
\geq 10^{-12}GeV$ on the noncommutativity parameter by setting the
noncommutative correction ratio $R$ to be of order one.
Although this limit on $1/\sqrt\theta$ is weak, we may also impose a
stronger limit on the magnitude of the noncommutativity parameter, by
measuring the difference of spin Hall conductivity at different
direction. However, there is no such experimental data available
yet.  We also have defined a new parameter
$\varsigma=\rho\theta$ which may be measurable experimentally as
well. The results in this paper suggest that high precision
measurements in quantum mechanical systems may be able to reveal the
noncommutativity of space.

\vskip 0.5cm \noindent\textbf{Acknowledgments}:  This work is
supported by the National Natural Science Foundation of China
(10965006).

\end{document}